\documentclass[american,aps,pre,reprint]{revtex4-1}
\usepackage[T1]{fontenc}
\usepackage[latin9]{inputenc}
\usepackage{amsmath}
\usepackage{graphicx}
\usepackage{esint}
\usepackage{babel}
\begin{document}

\title{Stable and unstable roots of ion temperature gradient driven mode
using curvature modified plasma dispersion functions}

\author{Ö. Gültekin$^{1}$, Ö. D. Gürcan$^{2,3}$}
\email{ozgur.gurcan@lpp.polytechnique.fr}

\selectlanguage{american}%

\affiliation{$^{1}$Department of Physics, Istanbul University, Istanbul, 34134,
Turkey}

\affiliation{$^{2}$CNRS, Laboratoire de Physique des Plasmas, Ecole Polytechnique,
Palaiseau}

\affiliation{$^{3}$ Sorbonne Universités, UPMC Univ Paris 06, Paris}
\begin{abstract}
Basic, local kinetic theory of ion temperature gradient driven (ITG)
mode, with adiabatic electrons is reconsidered. Standard unstable,
purely oscillating as well as damped solutions of the local dispersion
relation are obtained using a bracketing technique that uses the argument
principle. This method requires computing the plasma dielectric function
and its derivatives, which are implemented here using modified plasma
dispersion functions with curvature and their derivatives, and allows
bracketing/following the zeros of the plasma dielectric function which
corresponds to different roots of the ITG dispersion relation. We
provide an open source implementation of the derivatives of modified
plasma dispersion functions with curvature, which are used in this
formulation. Studying the local ITG dispersion, we find that near
the threshold of instability the unstable branch is rather asymmetric
with oscillating solutions towards lower wave numbers (i.e. drift
waves), and damped solutions toward higher wave numbers. This suggests
a process akin to inverse cascade by coupling to the oscillating branch
towards lower wave numbers may play a role in the nonlinear evolution
of the ITG, near the instability threshold. Also, using the algorithm,
the linear wave diffusion is estimated for the marginally stable ITG
mode.
\end{abstract}
\maketitle

\section{Introduction}

\subsection{Background}

Ion temperature gradient driven (ITG) mode was studied in great detail
over the years in light of its relevance for transport in magnetized
fusion devices\citep{coppi:67,horton:81,lee:86}. A basic formulation
of the kinetic ITG that has been studied in the past is a local, electrostatic
description based on the gyrokinetic equation\citep{catto:78,frieman:82,hahm:88}
for ions, with adiabatic electrons, where the linear problem boils
down to finding the roots of the plasma dielectric function $\varepsilon\left(\omega,\mathbf{k}\right)$
numerically.

Kinetic waves in electrostatic plasmas in general, can be described
using the so-called plasma dispersion function \citep{fried:1961}.
The cylindrical ITG mode for instance, can be formulated completely
in terms of plasma dispersion functions\citep{mattor:89}. The advantage
of such a formulation is that, the plasma dispersion function is linked
to the complex error function and there exists efficient methods for
its computation\citep{gautschi:70}.

Recently, a similar, numerically efficient reformulation of local
ITG in terms of curvature modified plasma dispersion functions was
proposed \citep{gurcan:14}, which is equivalent to the formulation
in Refs. \citep{kim:94,kuroda:98}. These functions, dubbed $I_{nm}\left(\zeta_{\alpha},\zeta_{\beta},b\right)$,
and defined for $Im\left[\zeta_{\alpha}\right]>0$ as: 
\begin{align}
 & I_{nm}\left(\zeta_{\alpha},\zeta_{\beta},b\right)\equiv\nonumber \\
 & \frac{2}{\sqrt{\pi}}\int_{0}^{\infty}dx_{\perp}\int_{-\infty}^{\infty}dx_{\parallel}\frac{x_{\perp}^{n}x_{\parallel}^{m}J_{0}^{2}\left(\sqrt{2b}x_{\perp}\right)e^{-x^{2}}}{\left(x_{\parallel}^{2}+\frac{x_{\perp}^{2}}{2}+\zeta_{\alpha}-\zeta_{\beta}x_{\parallel}\right)}\;\text{,}\label{eq:db_int}
\end{align}
can be written as a 1D integral of a combination of plasma dispersion
functions, instead of the two dimensional integral shown above. Note
that, since these functions have been formulated with built-in analytical
continuation, a dispersion relation, written with these functions,
can be used to describe oscillating and damped solutions as well as
unstable ones.

In this paper, we extend the space of curvature modified plasma dispersion
functions by including their derivatives {[}i.e. $J_{nm}\left(\zeta_{\alpha},\zeta_{\beta},b\right)\equiv-\frac{\partial}{\partial\zeta_{\alpha}}I_{nm}\left(\zeta_{\alpha},\zeta_{\beta},b\right)${]},
defined as
\begin{align}
 & J_{nm}\left(\zeta_{\alpha},\zeta_{\beta},b\right)\equiv\nonumber \\
 & \frac{2}{\sqrt{\pi}}\int_{0}^{\infty}dx_{\perp}\int_{-\infty}^{\infty}dx_{\parallel}\frac{x_{\perp}^{n}x_{\parallel}^{m}J_{0}^{2}\left(\sqrt{2b}x_{\perp}\right)e^{-x^{2}}}{\left(x_{\parallel}^{2}+\frac{x_{\perp}^{2}}{2}+\zeta_{\alpha}-\zeta_{\beta}x_{\parallel}\right)^{2}}\;\text{,}\label{eq:jnm1}
\end{align}
and use these functions in order to compute the derivatives of the
plasma dielectric function $\varepsilon\left(\omega,\mathbf{k}\right)$
with respect to the angular frequency $\omega$. This allows the use
of a root finding alogrithm based on the argument principle as discussed
for instance in Ref.\citep{johnson:09} (similar to the method used
in the quasi-linear solver QualiKiz\citep{bourdelle:07} and detailed
in Ref. \citep{davies:86}), which allows us to obtain both unstable
and stable roots.

Notice that in most cases, the stable roots are considered to have
a negligible effect on transport and are therefore ignored. However,
from a simple quasi-linear theory (QLT) point of view, this is clearly
not permissible, since as the nonlinear interactions appear, so does
the transfer of energy to stable or damped modes. However, from a
renormalized QLT perspective -\emph{à la} Balescu \citep{balescu:book:anom},
which is actually how the use of QLT to estimate transport is really
justified- it is unclear whether one may use a single dominant (but
renormalized) mode, or one still has to consider a coupling of a number
of stable and unstable modes (even if each one of those modes are
modified due to nonlinear effects, via mechanisms such as eddy damping).
This may be crucial in particular if after renormalization, the most
unstable (or the least damped) mode for a given wave-vector becomes
subdominant to a previously subdominant mode.

The rest of the paper is organized as follows. In section II, the
local ITG dispersion relation is recalled using curvature modified
plasma dispersion functions, $I_{nm}$'s. Then, in subsection b),
derivatives of the curvature modified dispersion functions are defined
as $J_{nm}$'s, and the derivative of the plasma dielectric tensor
is written in terms of $J_{nm}$'s. In Section III, methods and examples,
first an efficient and accurate method for finding and tracing the
roots of the dispersion relation is introduced in subsection a), and
then an example of linear wave diffusion of an unstable mode into
linearly stable region is considered and the diffusion coefficient
is estimated. Section IV is results and conclusion. 

\begin{widetext}

\section{Formulation}

\subsection{Linear Dispersion Using $I_{nm}$'s:}

A basic description of local kinetic ITG in the electrostatic limit,
with adiabatic electrons is based on the gyrokinetic equation \citep{frieman:82,lee:83,hahm:88}
for the non-adiabatic part of the fluctuating distribution function
for the ions:
\begin{align}
\frac{\partial}{\partial t}\delta g+ & \left[v_{\parallel}\frac{\mathbf{B}^{*}}{B}+\frac{\mu}{eB}\hat{\mathbf{b}}\times\nabla B\right]\cdot\nabla\delta g=\frac{e}{T_{i}}F_{0}\frac{\partial}{\partial t}\left\langle \delta\Phi\right\rangle -F_{0}\frac{\hat{\mathbf{b}}}{B}\times\nabla\left\langle \delta\Phi\right\rangle \cdot\left[\frac{1}{n}\nabla n+\left(\frac{E}{T}-\frac{3}{2}\right)\frac{1}{T}\nabla T\right]\;\text{.}\label{eq:deltageqn-1}
\end{align}
This is then complemented by the quasi-neutrality relation ($n_{e}=n_{i}$),
with adiabatic electrons:
\begin{equation}
\frac{e}{T_{e}}\Phi=-\frac{e\Phi}{T_{i}}+\int J_{0}\delta gd^{3}v\;\text{.}\label{eq:qn}
\end{equation}
Taking the Laplace-Fourier transform of (\ref{eq:deltageqn-1}) in
the form $\delta g_{\mathbf{k},\omega}\left(\mathbf{v}\right)=\int e^{-i\omega t+i\mathbf{k}\cdot\mathbf{x}}\delta g\left(\mathbf{x},\mathbf{v},t\right)$
and solving for $\delta g_{\mathbf{k},\omega}$ and substituting the
result into (\ref{eq:qn}), we obtain the dispersion relation in the
form:

\begin{equation}
\varepsilon\left(\omega,\mathbf{k}\right)\equiv1+\frac{1}{\tau}-\left[\frac{1}{\sqrt{2\pi}v_{ti}^{3}}\int\frac{\left(\omega-\omega_{*Ti}\left(v\right)\right)J_{0}\left(\frac{v_{\perp}k_{\perp}}{\Omega_{i}}\right)^{2}}{\left(\omega-v_{\parallel}k_{\parallel}-\omega_{Di}\frac{1}{2}\left(\frac{v_{\parallel}^{2}}{v_{ti}^{2}}+\frac{v_{\perp}^{2}}{2v_{ti}^{2}}\right)\right)}e^{-\frac{v^{2}}{2v_{ti}^{2}}}v_{\perp}dv_{\perp}dv_{\parallel}\right]=0\;\text{,}\label{eq:drel-1}
\end{equation}
\end{widetext}where $\varepsilon\left(\omega,\mathbf{k}\right)$
is the plasma dielectric function, 
\[
\omega_{*Ti}\left(v\right)\equiv\omega_{*i}\left[1+\left(\frac{v^{2}}{2v_{ti}^{2}}-\frac{3}{2}\right)\eta_{i}\right]\;\mbox{,}
\]
and $\omega_{Di}=2\frac{L_{n}}{R}\omega_{*i}$. Using $\omega/\left|k_{y}\right|\rightarrow\omega$,
and $\omega_{D}/\left|k_{y}\right|\rightarrow\omega_{D}$, the dispersion
relation (\ref{eq:drel-1}) can be written as:

\begin{align}
\varepsilon\left(\omega,\mathbf{k}\right)\equiv & 1+\frac{1}{\tau}+\frac{1}{\omega_{Di}}\bigg(I_{10}\left[\omega+\left(1-\frac{3}{2}\eta_{i}\right)\right]\nonumber \\
 & +\left(I_{30}+I_{12}\right)\eta_{i}\bigg)=0\label{eq:eps_pdf}
\end{align}
where $I_{nm}\equiv I_{nm}\left(-\frac{\omega}{\omega_{Di}},-\frac{\sqrt{2}k_{\parallel}}{\omega_{Di}k_{y}},b\right)$.
The advantage of this particular form is that the explicit scaling
of $\omega$ with $k_{y}$ is removed, so that we can define a region
in $\omega$ space to search for roots, and do not need to scale it
with $k_{y}$. 

As discussed in detail in Ref. \citealp{gurcan:14}, the $I_{nm}$'s
can be written as a single integral:

\begin{align*}
 & I_{nm}\left(\zeta_{\alpha},\zeta_{\beta},b\right)=\\
 & \int_{0}^{\infty}s^{\frac{n-1}{2}}G_{m}\left(z_{1}\left(s\right),z_{2}\left(s\right)\right)J_{0}\left(\sqrt{2bs}\right)^{2}e^{-s}ds\;\mbox{,}\\
 & \quad(Im\left[\zeta_{\alpha}\right]>0)
\end{align*}
using the straightforward multi-variable generalization of the standard
plasma dispersion function:
\begin{equation}
G_{m}\left(z_{1},z_{2},\cdots,z_{n}\right)\equiv\frac{1}{\sqrt{\pi}}\int_{-\infty}^{\infty}\frac{x^{m}e^{-x^{2}}}{\prod_{i=1}^{n}\left(x-z_{i}\right)}dx\label{eq:def_gen}
\end{equation}
with
\begin{equation}
z_{1,2}\left(s\right)=\frac{1}{2}\left(\zeta_{\beta}\pm\sqrt{\zeta_{\beta}^{2}-2\left(s+2\zeta_{\alpha}\right)}\right)\;\mbox{.}\label{eq:z12}
\end{equation}
Note that, using Eqns. 4 and 6 of Ref. \citealp{gurcan:14}, we can
write the $G_{m}\left(z_{1},z_{2}\right)$ in terms of the standard
plasma dispersion function as:

\begin{align}
G_{m}\left(z_{1},z_{2}\right) & =\frac{1}{\sqrt{\pi}\left(z_{1}-z_{2}\right)}\bigg[z_{1}^{m}Z_{0}\left(z_{1}\right)-z_{2}^{m}Z_{0}\left(z_{2}\right)\nonumber \\
 & +\sum_{k=2}^{m}\left(z_{1}^{k-1}-z_{2}^{k-1}\right)\Gamma\left(\frac{m-k+1}{2}\right)\bigg]\;,\label{eq:gm}
\end{align}
which was then implemented using a 16 coefficient Weideman method
\citep{weideman:94}, in the form of an open source fortran library
{[}\url{http://github.com/gurcani/zpdgen}{]} with a python interface.

As discussed in Refs. \citealp{kim:94} and \citealp{kuroda:98}.
In addition to the integral in (\ref{eq:eps_pdf}), the analytical
continuation requires adding a residue contribution, which can be
computed as \begin{widetext}
\begin{align*}
\Delta I_{nm}\left(\zeta_{\alpha},\zeta_{\beta},b\right)=-i\sqrt{\pi}2^{\frac{\left(n+3\right)}{2}}w^{\frac{n}{2}}\int_{-1}^{1}d\mu & \left(1-\mu^{2}\right)^{\frac{\left(n-1\right)}{2}}\left(\mu\sqrt{w}+\frac{\zeta_{\beta}}{2}\right)^{m}\\
 & J_{0}^{2}\left(2\sqrt{b\left(1-\mu^{2}\right)w}\right)e^{-2\left(1-\mu^{2}\right)w-\left(\mu\sqrt{w}+\frac{\zeta_{\beta}}{2}\right)^{2}}\times\begin{cases}
0 & \zeta_{\alpha i}>0\quad\mbox{or }w_{r}<0\\
\frac{1}{2} & \zeta_{\alpha i}=0\quad\mbox{and }w_{r}>0\\
1 & \zeta_{\alpha i}<0\quad\mbox{and }w_{r}>0
\end{cases}
\end{align*}
\end{widetext}where $w=\frac{\zeta_{\beta}^{2}}{4}-\zeta_{\alpha}$
and $\zeta_{\alpha i}=\text{Im}\left(\zeta_{\alpha}\right)$ and $w_{r}=\text{Re}\left[w\right]$.
With this, $I_{nm}=I_{nm}^{'}+\Delta I_{nm}$ {[}where $I_{nm}^{'}$
is the integral in (\ref{eq:eps_pdf}){]} is defined everywhere on
the complex plane.

\subsection{Derivatives of $I_{nm}$'s}

Similarly the derivatives as defined by the relation
\[
J_{nm}\left(\zeta_{\alpha},\zeta_{\beta},b\right)\equiv-\frac{\partial}{\partial\zeta_{\alpha}}I_{nm}\left(\zeta_{\alpha},\zeta_{\beta},b\right)
\]
 can be written using 
\begin{align}
 & J_{nm}\left(\zeta_{\alpha},\zeta_{\beta},b\right)=\nonumber \\
 & \int_{0}^{\infty}ds\left[s^{\frac{n-1}{2}}G_{m}\left(z_{1},z_{2},z_{1},z_{2}\right)J_{0}\left(\sqrt{2bs}\right)^{2}e^{-s}\right]\;\mbox{,}\nonumber \\
 & \quad(Im\left[\zeta_{\alpha}\right]>0)\label{eq:jnm2}
\end{align}
with repeating variables $z_{1}=z_{1}\left(s\right)$ and $z_{2}=z_{2}\left(s\right)$
as given in (\ref{eq:z12}). Since the $\zeta_{\alpha}$ dependence
is through $z_{1}$ and $z_{2}$, we can use (\ref{eq:z12}) to compute
the derivatives , acting $\frac{d}{d\zeta_{\alpha}}=\frac{1}{\left(z_{1}-z_{2}\right)}\left(\frac{d}{dz_{2}}-\frac{d}{dz_{1}}\right)$
on (\ref{eq:gm}), in order to obtain:
\begin{figure}
\begin{centering}
\includegraphics[width=0.98\columnwidth]{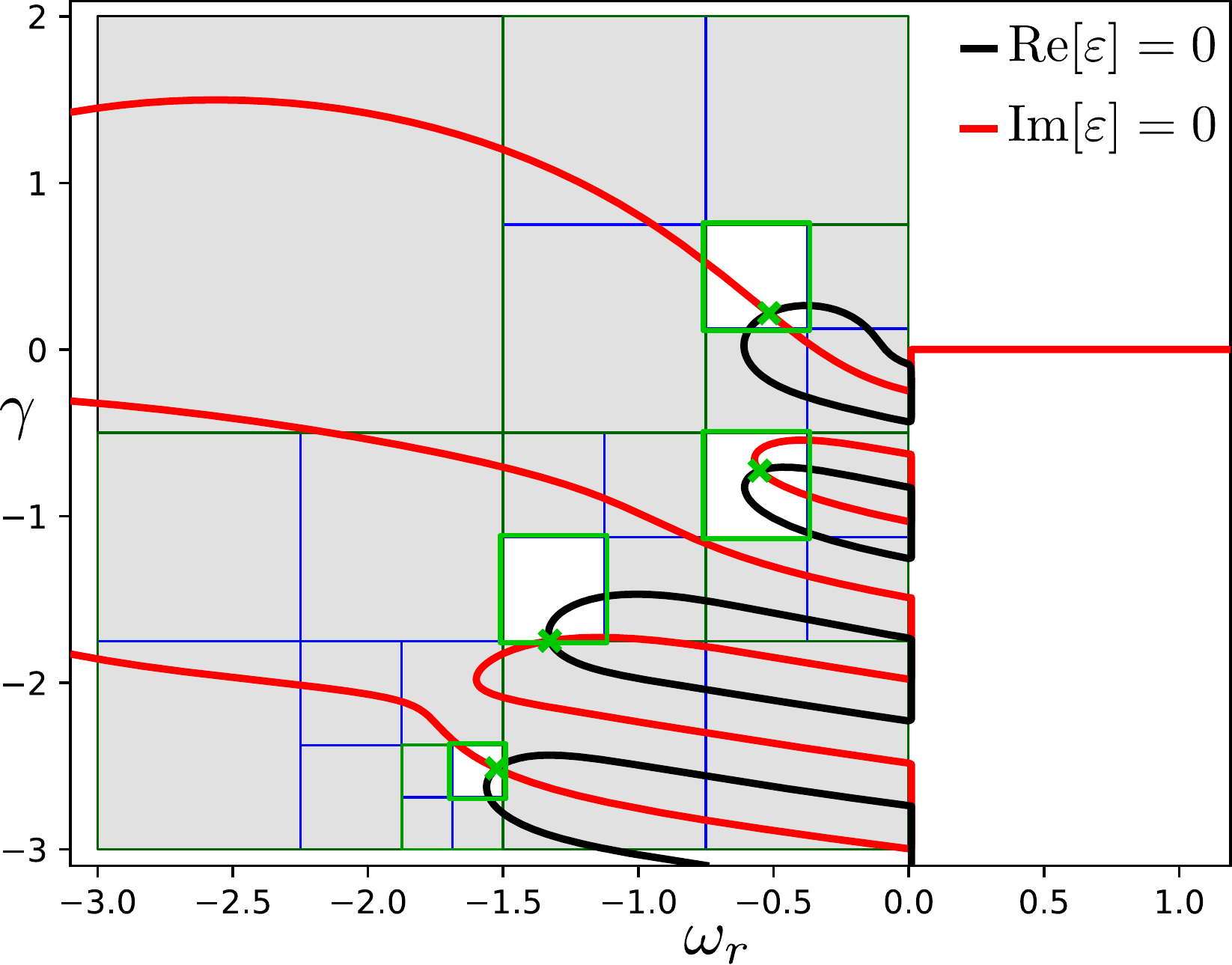}
\par\end{centering}
\caption{\label{fig:rects}The way the bracketing algorithm isolates the roots
of the plasma dielectric function (\ref{eq:eps_pdf}) to a desired
rectangle size. Shaded rectangles contain no roots and so are immediately
abandoned. The roots, which are depicted by x's are found using a
least square optimization, where the midpoint of the final rectangle
is used as the initial guess, and the rectangle itself is used as
a boundary. The case that is shown here is $k_{y}=0.8$, which is
usually used as the reference $k_{y}$.}
\end{figure}
\begin{widetext}
\begin{align}
G_{m}\left(z_{1},z_{2},z_{1},z_{2}\right) & =-\frac{d}{d\zeta_{\alpha}}G_{m}\left(z_{1},z_{2}\right)=\frac{1}{\left(z_{1}-z_{2}\right)^{2}}\bigg\{\frac{1}{\sqrt{\pi}}\sum_{k=2}^{m}\left[\left(k-1\right)\left(z_{1}^{k-2}+z_{2}^{k-2}\right)-\frac{2\left(z_{1}^{k-1}-z_{2}^{k-1}\right)}{\left(z_{1}-z_{2}\right)}\right]\Gamma\left(\frac{m-k+1}{2}\right)\nonumber \\
 & -2\left(z_{1}^{m}+z_{2}^{m}\right)+Z_{0}\left(z_{1}\right)z_{1}^{m-1}\left(m-2z_{1}^{2}-\frac{2z_{1}}{\left(z_{1}-z_{2}\right)}\right)+Z_{0}\left(z_{2}\right)z_{2}^{m-1}\left(m-2z_{2}^{2}+\frac{2z_{2}}{\left(z_{1}-z_{2}\right)}\right)\bigg\}\text{\;.}\label{eq:gm2}
\end{align}
We also have to compute the derivatives of the residue contribution
$\Delta I_{nm}$, which we dub $\Delta J_{nm}$ (note that, this is
derivative of the residue and not the residue of the derivative),
and can be written as: 
\begin{align}
\Delta J_{nm}= & -i\sqrt{\pi}2^{\frac{n+3}{2}}\int_{-1}^{1}\bigg\{\left(\frac{n}{2w}+\frac{\mu m}{2\mu w+\sqrt{w}\zeta_{\beta}}-2+\mu^{2}-\frac{\zeta_{\beta}}{2}\frac{\mu}{\sqrt{w}}\right)J_{0}^{2}\left(2\sqrt{b\left(1-\mu^{2}\right)w}\right)\nonumber \\
 & -2\sqrt{\frac{b\left(1-\mu^{2}\right)}{w}}J_{0}\left(2\sqrt{b\left(1-\mu^{2}\right)w}\right)J_{1}\left(2\sqrt{b\left(1-\mu^{2}\right)w}\right)\bigg\}\nonumber \\
 & w^{n/2}\left(1-\mu^{2}\right)^{\frac{n-1}{2}}\left(\mu\sqrt{w}+\frac{\zeta_{\beta}}{2}\right)^{m}e^{-2\left(1-\mu^{2}\right)w-\left(\mu\sqrt{w}+\frac{\zeta_{\beta}}{2}\right)^{2}}d\mu\times\begin{cases}
0 & \zeta_{\alpha i}>0\quad\mbox{or }w_{r}<0\\
\frac{1}{2} & \zeta_{\alpha i}=0\quad\mbox{and }w_{r}>0\\
1 & \zeta_{\alpha i}<0\quad\mbox{and }w_{r}>0
\end{cases}\label{eq:djnm}
\end{align}
\end{widetext}where we used the definition $\Delta J_{nm}=-\frac{d}{d\zeta_{\alpha}}\Delta I_{nm}=\frac{d}{dw}\Delta I_{nm}$.
Finally, $J_{nm}=J_{nm}^{'}+\Delta J_{nm}$ where $J_{nm}^{'}$ is
the integral given in (\ref{eq:jnm2}) with (\ref{eq:gm2}).

Using these $J_{nm}$ functions, which denote derivatives of the curvature
modified plasma dispersion functions with respect to the first variable,
the derivative of the plasma dielectric function can be written as:
\begin{align}
\frac{\partial}{\partial\omega}\varepsilon\left(\omega,\mathbf{k}\right)\equiv\frac{1}{\omega_{Di}}I_{10} & +\frac{1}{\omega_{D}^{2}}\bigg(J_{10}\left[\omega+\left(1-\frac{3}{2}\eta_{i}\right)\right]\nonumber \\
 & +\left(J_{30}+J_{12}\right)\eta_{i}\bigg)\;\text{.}\label{eq:deps}
\end{align}
where $J_{nm}\equiv J_{nm}\left(-\frac{\omega}{\omega_{Di}},-\frac{\sqrt{2}k_{\parallel}}{\omega_{Di}k_{y}},b\right)$,
and $\omega$ and $\omega_{D}$ are normalized to $\left|k_{y}\right|$
for convenience.

\section{Methods and Examples}

\subsection{Finding and tracking stable and unstable solutions}

Fixing the values of plasma parameters such as $\eta_{i}$, $R/L_{n}$
and $\tau$, we can solve (\ref{eq:eps_pdf}) for $\omega$, for a
given $\mathbf{k}$. In practice we fix $k_{\parallel}$ and $k_{x}$
and consider $\omega$ as a function of $k_{y}$. While there are
many different ways of achieving this numerically, we have developed
a simple algorithm for bracketing, solving and then tracing each root
of the solution. Generally we pick a reference $k_{y}$ value, where
we think the roots are reasonably distinct (choosing this reference
$k_{y}$ may require trial and error). Then we use an algorithm very
similar to the one outlined in Ref. \citealp{johnson:09} in order
to bracket each solution as shown in Fig. \ref{fig:rects}, with an
initial rectangle that covers only the $\omega_{r}<0$ part of the
complex plane avoiding the line $\omega_{r}=0$, where there is a
branch cut. In fact a desired number of roots (i.e. $N_{r}$) are
specified so that the algoritm repeats itself with larger and larger
rectangles (always avoiding the branch cut) until the desired number
of roots fall within the rectangle. This gives us $N_{r}$ rectangles
with a root in each one. Then, a basic least square optimization is
used to locate the exact root within each rectangle. Note that a small
buffer is added around the boundary of the rectangle in order to succeed
in cases where the point falls exactly on the boundary of the rectangle
(e.g. third root from above in Fig. \ref{fig:rects}). 

When the $k_{y}$ is varied, a new rectangle is defined for each $N_{r}$
root, using the solutions from one of the previous steps (i.e. nearest
$k_{y}$, for which $\omega$ have already been computed), with a
predefined rectangle size (i.e. if $k_{y}$ resolution is high enough,
the rectangle sizes can be very small and there are virtually no intersections),
and the least square optimization is used again to find the new solutions
in each rectangle. This allows us to trace curves in $\omega=\omega\left(k_{y}\right)$,
which help distinguish different roots. Note that tracking $\omega$
as a function of $k_{y}$ instead of repeating the bracketing step
each time, saves a huge amount of computation time. Such an approach
would be useful also in quasilinear transport modelling geared towards
speed\citep{bourdelle:07}. 
\begin{figure}
\begin{centering}
\includegraphics[width=0.98\columnwidth]{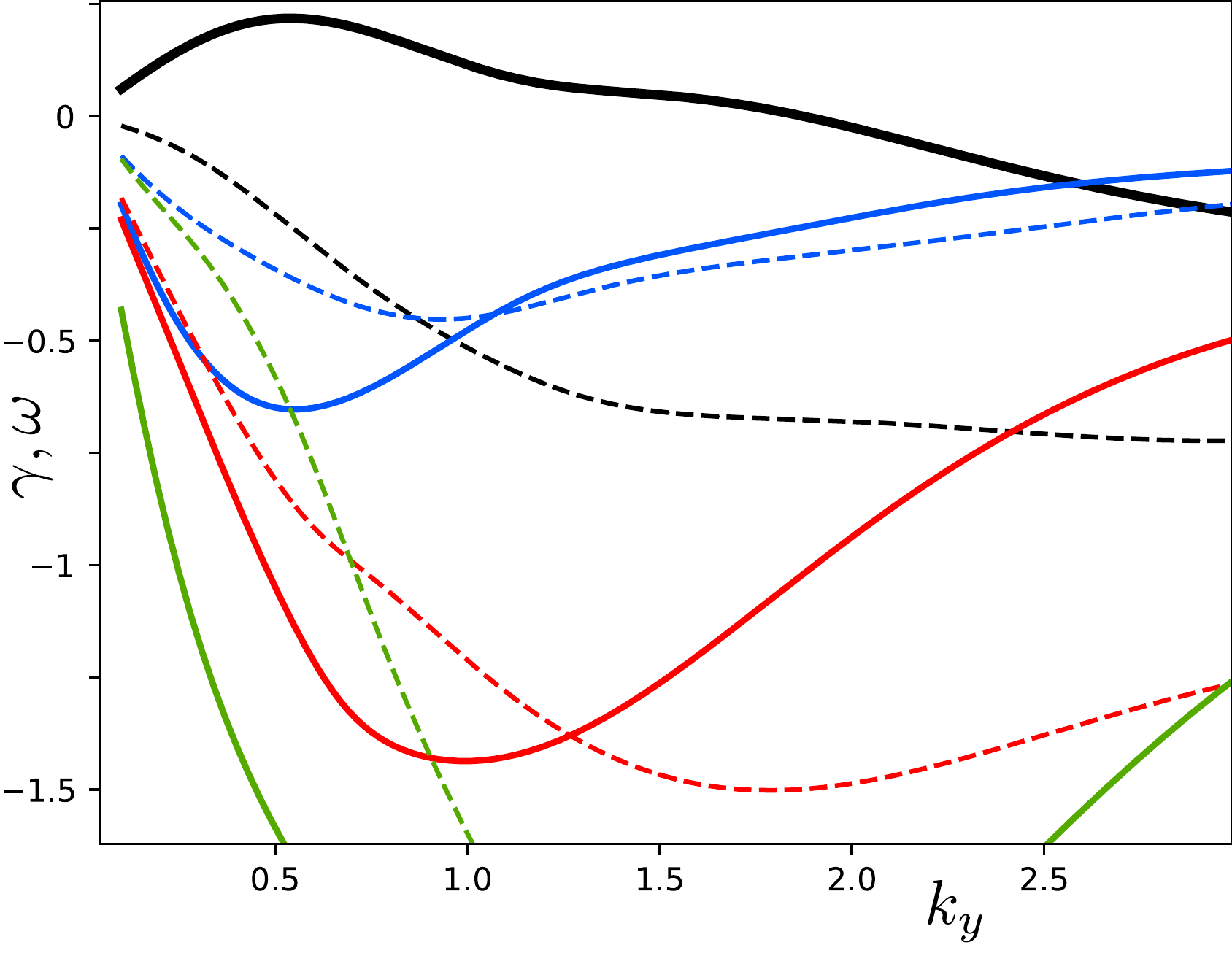}
\par\end{centering}
\caption{\label{fig:gamma}Growth rates $\gamma$ (solid lines), and frequencies
$\omega$ (dashed lines) as functions of $k_{y}$, for the first four
roots of the local, kinetic ITG dispersion relation as defined in
(\ref{eq:eps_pdf}), where each color denotes a seperate root. Note
that around $k_{y}=2.5$, the second root becomes less damped than
the unstable branch.}
\end{figure}

In any case, bracketing is necessary in order to isolate the different
roots of (\ref{eq:eps_pdf}). Since the algorithm relies on the argument
principle
\[
\oint_{C}\frac{\frac{\partial}{\partial\omega}\varepsilon\left(\omega,\mathbf{k}\right)}{\varepsilon\left(\omega,\mathbf{k}\right)}d\omega=2\pi i\left(N-P\right)
\]
where $N$ and $P$ are the number of poles and zeros in the closed
contour defined by$C$, we use (\ref{eq:deps}) in order to compute
the derivative of the plasma dielectric function analytically. 
\begin{figure}
\begin{centering}
\includegraphics[width=0.98\columnwidth]{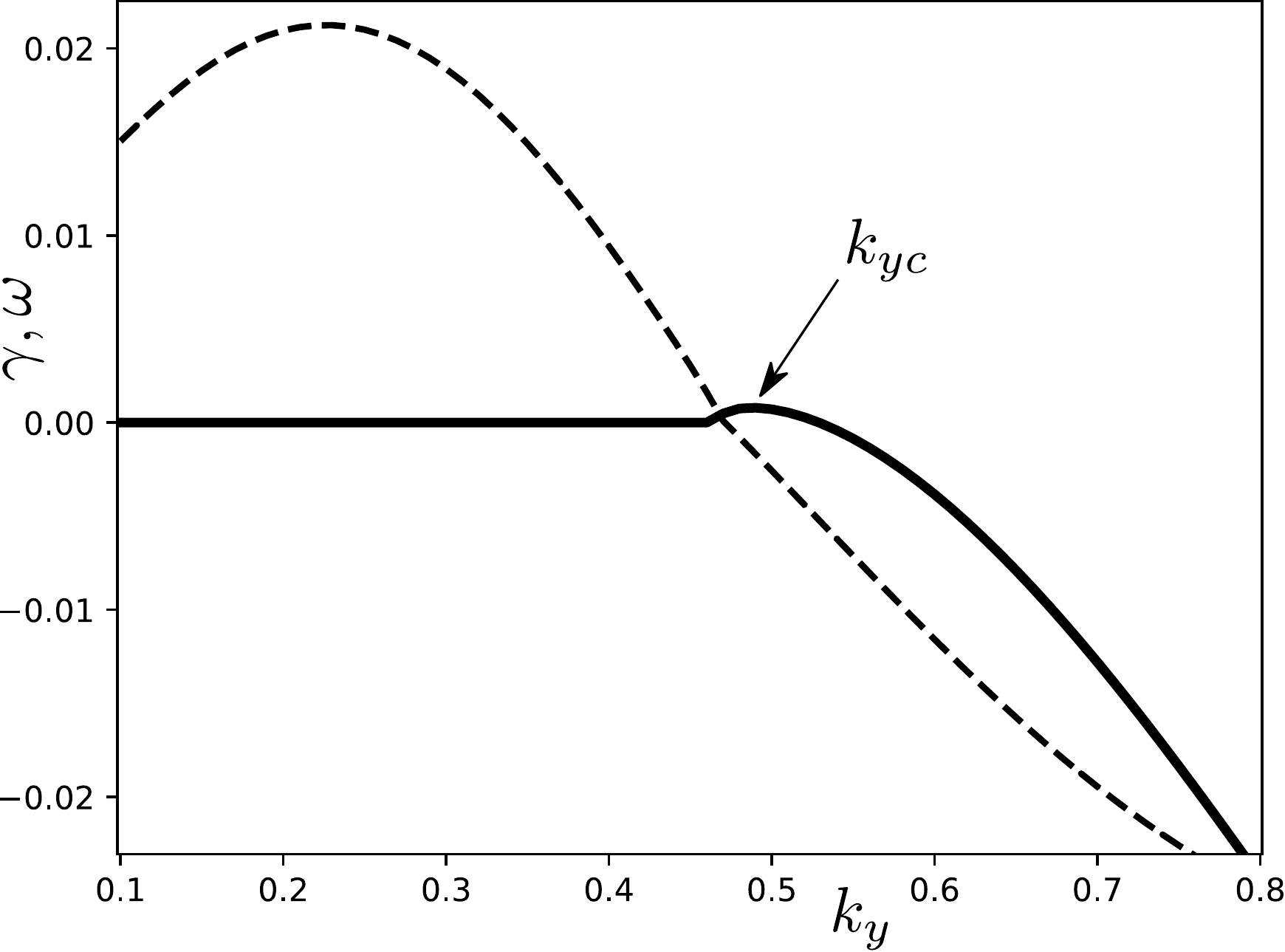}
\par\end{centering}
\caption{\label{fig:gamma2}Growth rate $\gamma$ (solid line), and frequency
$\omega$ (dashed lines) as functions of $k_{y}$, for the dominant
root of the ITG dispersion relation near the threshold of instability
(i.e. $\eta_{i}=0.68$). Note that the damped modes in this case,
are strongly damped (i.e. $\gamma_{d}<-0.2$) as compared to the unstable
mode.}
\end{figure}

Using this method, the growth (and damping) rates as well as frequencies
for the reference shot studied in Ref. \citealp{kim:94} (i.e. $\eta_{i}=2.5$,
$L_{n}/R=0.2$, $\tau=1.0$ and $k_{\parallel}=0.01$) is shown in
Fig. \ref{fig:gamma}. It is remarkable that for these parameters,
around $k_{y}=2.5$, the second root becomes less damped than the
unstable branch. Since trapped electron physics is ignored, the second
root never actually becomes unstable.

Another interesting observation about the nature of the roots of this
particular limit of the gyrokinetic equation, is that near the instability
threshold (slightly above, or slightly below), one observes a region
to the left (in $k_{y}$ space) of the linearly most unstable (or
the least damped) mode, where the solution becomes a propagating wave
in the electron diamagnetic direction (i.e. $\omega>0$) as seen in
Fig. \ref{fig:gamma2}. This is the drift wave (DW) branch, that is
modified due to the weak ion temperature gradient. This is not a surprise,
since the equations considered in this paper should recover the drift
wave limit as the ITG drive disappears.
\begin{figure*}
\begin{centering}
\includegraphics[width=0.98\textwidth]{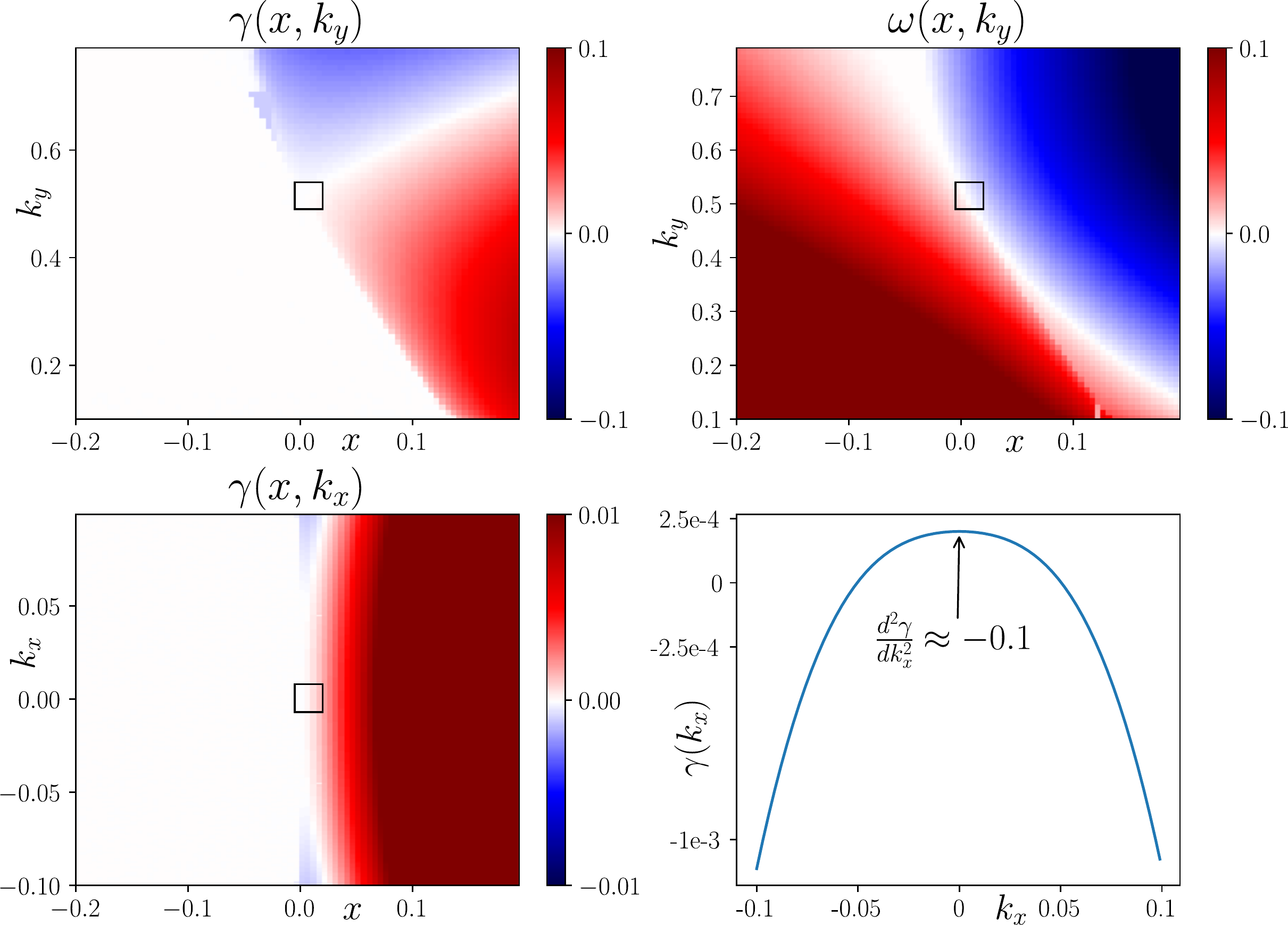}
\par\end{centering}
\caption{\label{fig:Profiles}Profiles of a) the growth rate $\gamma_{k}$
as a function of position ($x$-axis) and $k_{y}$ (y-axis) at $k_{x}=0$,
b) the real frequency as a function of position ($x$-axis) and $k_{y}$
(y-axis) at $k_{x}=0$, c) the growth rate $\gamma_{k}$ as a function
of position ($x$-axis) and $k_{x}$ (y-axis) at $k_{y}=k_{yc}$,
where $k_{yc}$ is the most unstable wavenumber at $x=k_{x}=0$, d)
the growth rate $\gamma_{k}$ as a function of $k_{x}$ at $k_{y}=k_{yc}$
and $x=0$. The box shows the marginal point. As can be seen in a)
and b), the growth rate $\gamma=0$ to the left of the marginal point
(i.e. for which $\eta_{i}<\eta_{ic}$), where the frequency is positive
(i.e. electron diamagnetic direction).}
\end{figure*}

\subsection{Example: Linear diffusion near marginality}

For a given set of plasma parameters, $\eta_{i}$ determines the stability
of the ITG mode. Considering $\eta_{i}$ as a function of $x$ , for
instance of the form $\eta_{i}\left(x\right)=\eta_{ic}+\delta\eta_{i}\left[\frac{x-x_{0}}{x_{1}-x_{0}}\right]$,
we can define a resonable description of an unstable region next to
a stable region. The issue of turbulence spreading into the unstable
region is a complex one and is out of the scope of the current paper.
Here we discuss how a monochromatic wave, propagating mainly in the
$y$ direction can diffuse in the radial direction due to $\partial^{2}\gamma_{k}/\partial k_{x}^{2}$
being finite and negative. Consider the evolution of the amplitude
of ITG mode near its stability boundary. Close enough to the marginal
stability, only a single mode will be linearly unstable. We can write
the general two scale evolution equation for the amplitude of that
mode in the form
\begin{equation}
\left(\partial_{t}+v_{gi}\partial_{i}\right)I_{k}-2\gamma_{k_{x},k_{y}}I_{k}-D_{ij}\partial_{ij}I_{k}+\gamma_{n\ell}I_{k}^{2}=0\label{eq:landau}
\end{equation}
Here $I=\left|\Phi_{k}\right|^{2}$ is the intensity of the most unstable
mode, $v_{gi}=\partial\omega_{k}/\partial k_{i}$, $D_{ij}=-\partial^{2}\gamma_{k_{x},k_{y}}/\partial k_{i}\partial k_{j}$
and $\gamma_{n\ell}$ is the nonlinear damping via mode coupling or
coupling to large scale flows, whose origin is, again, out of the
scope here. Nonetheless, the local mixing length estimate would suggest
$\gamma_{n\ell}\sim2k_{\perp}^{2}$. Note that the most unstable mode
has $\frac{\partial}{\partial k_{x}}\gamma=\frac{\partial}{\partial k_{y}}\gamma=0$
, by definition and has $\omega_{k}\approx0$ for the ITG mode near
marginality. In addition it is also true that $\partial\omega/\partial k_{x}\ll\partial\omega/\partial k_{y}$,
near the stability boundary.

Eqn. (\ref{eq:landau}) is a linear Fisher-Kolmogorov equation\citep{fisher:37}
similar to the one discussed in the study of formation of subcritical
turbulence fronts\citep{pomeau:86}. Moving to the group velocity
frame in the $y$ direction, and considering mainly the diffusion
in the $x$ direction, we get:
\begin{equation}
\partial_{t}I-2\gamma_{k}\left(x\right)I-D_{xx}\partial_{xx}I+\gamma_{n\ell}I^{2}=0\label{eq:landau2}
\end{equation}
 Using $\theta=\frac{k_{x}}{\hat{s}k_{y}}$, with $\omega_{D}\rightarrow\omega_{D}\left(\cos\theta+\hat{s}\theta\sin\theta\right)$
and $b=k_{\perp}^{2}=k_{y}^{2}+k_{x}^{2}$ in (\ref{eq:drel-1}),
in order to get the $k_{x}$ dependence of the growth rate, we can
obtain the growth rate and frequency as a function of $k_{x}$, $x$
and $k_{y}$ as show in Fig. \ref{fig:Profiles}. This allows us to
compute a linear diffusion coefficient via $D_{xx}=-\partial^{2}\gamma/\partial k_{x}^{2}$,
which can be estimated to be around $D_{xx}\approx0.1$, near the
the marginal point. More generally, the methodology that we have developed
above allows us to determine the coefficients of (\ref{eq:landau2})
- except $\gamma_{n\ell}$.

\section{Results and Conclusion}

The method outlined in this paper allows us to solve the local linear
gyrokinetic equation with adiabatic electrons and background density
and ion temperature gradients as in (\ref{eq:deltageqn-1}-\ref{eq:qn})
for stable and unstable roots using generalized plasma dispersion
functions as seen in Fig. \ref{fig:gamma}. It can be used to study
the behaviour of the ITG mode near $\eta_{i}=\eta_{ic}$ for the instability
(i.e. $\eta_{ic}=2/3$ for small enough $R/L_{n}$), where the subcritical
solution becomes a propagating wave in the electron diamagnetic direction
(i.e. $\omega>0$) as seen in Fig. \ref{fig:gamma2}. This is the
drift wave (DW) branch, that is modified due to the weak ion temperature
gradient.

The existence of a drift wave with $\gamma=0$, has important implications
for subcritical turbulence, especially when one considers a stable
region next to an ITG unstable region. In such a scenario, the ITG
that is generated at the unstable region with higher wavenumbers (say
around $k_{y}\sim0.3-0.5$) can couple to drift waves in the stable
region, which has the nice property of having $\gamma=0$ (rather
than negative) even when the $\eta_{i}$ is below critical. In this
case the wave diffusion (as discussed above due to $d^{2}\gamma/dk_{x}^{2}$)
or nonlinear spreading due to turbulent diffusion of broadband turbulence\citep{gurcan:05}
is easier, since the subcritical region does not act as a sink. 

It is also worth discussing the possibility of asymmetry in three
wave couplings near the threshold of the ITG instability. In the standard
picture of triadic interactions, the middle wave-number of a triad
gives its energy to larger and smaller wavenumbers, this normally
contributes equal amount to forward and backward cascades. However
since the higher $k_{y}$ 's are damped but lower $k_{y}$'s are not,
the energy would travel towards the drift wave branch naturally. Notice
that near marginal stability, the frequencies are such that it is
easy to satisfy the resonance conditions with a positive frequency
for the drift wave, and the negative frequency for the damped higher-$k_{y}$
ITG mode with $\omega\approx0$ for the pump. This may explain how
the free energy can be transfered to low $k_{y}$ in a process similar
to -but intrinsically different from- the inverse cascade.

The approach, developed in this paper, can be extended to a renormalized
version of the plasma dielectric function\citep{dupree:67}. However,
since both the careful implementation and the detailed analysis of
the physics results of such a formulation requires dedicated effort,
we leave this to future studies.

\end{document}